\documentclass[sigconf]{acmart}

\AtBeginDocument{%
 \providecommand\BibTeX{{%
   \normalfont B\kern-0.5em{\scshape i\kern-0.25em b}\kern-0.8em\TeX}}}
    
\usepackage{xspace}
\newcommand{\ie}{\textit{i.e.}}
\newcommand{\eg}{\textit{e.g.}}
\newcommand{\etal}{\textit{et al. \xspace}}
\usepackage{array}
\newcolumntype{L}{>{\arraybackslash}m{16cm}}
\newcolumntype{C}[1]{>{\centering\let\newline\\arraybackslash\hspace{0pt}}m{#1}}
\newcolumntype{R}[1]{>{\raggedleft\let\newline\\arraybackslash\hspace{0pt}}m{#1}}
\usepackage{adjustbox}
\usepackage{subcaption}
\usepackage{textcomp}
\usepackage[numbered]{bookmark}
\usepackage{soul}
\usepackage{booktabs}
\usepackage{multirow}
\usepackage{float}
\usepackage{url}
\usepackage{tcolorbox}
\usepackage{graphicx}
\usepackage{textcomp}
\usepackage{xcolor}
\usepackage{url}
\usepackage{tabularx} 
\usepackage{balance}
\usepackage{stfloats}
\usepackage[justification=centering]{caption}
\usepackage{lineno}
\usepackage[title]{appendix}
\usepackage{dirtytalk}
\usepackage{amsmath}
\usepackage{colortbl}

\newcommand{\RQone}{How do refactoring reviews compare to non-refactoring reviews in terms of code review efforts?\xspace}
\newcommand{\RQtwo}{What are the criteria that mostly associated with refactoring review decision?\xspace}

\begin{document}


\title{
{Code Review Practices for Refactoring Changes}: \\An Empirical Study on OpenStack} 

\author{Eman Abdullah AlOmar}
\affiliation{
    \institution{Stevens Institute of Technology}
    \city{Hoboken, New Jersey}
    \country{USA}
}
\email{ealomar@stevens.edu}

\author{Moataz Chouchen}
\affiliation{
    \institution{ETS Montreal, University of Quebec}
    \city{Montreal, Quebec}
    \country{Canada}
}
\email{moataz.chouchen.1@ens.etsmtl.ca}

\author{Mohamed Wiem Mkaouer}
\affiliation{
    \institution{Rochester Institute of Technology}
    \city{Rochester, New York}
    \country{USA}
}
\email{mwmvse@rit.edu}

\author{Ali Ouni}
\affiliation{
    \institution{ETS Montreal, University of Quebec}
    \city{Montreal, Quebec}
    \country{Canada}
}
\email{ali.ouni@etsmtl.ca}

\renewcommand{\shortauthors}{AlOmar et al.}

\renewcommand{\shortauthors}{AlOmar et al.}

\begin{abstract}
Modern code review is a widely used technique employed in both industrial and open-source projects to improve software quality, share knowledge, and ensure adherence to coding standards and guidelines. During code review, developers may discuss refactoring activities before merging code changes in the code base. To date, code review has been extensively studied to explore its general challenges, best practices and outcomes, and socio-technical aspects. However, little is known about how refactoring is being reviewed and what developers care about when they review refactored code. Hence, in this work, we present a quantitative and qualitative study  to understand what are the main criteria developers rely on to develop a decision about accepting or rejecting a submitted refactored code, and what makes this process challenging. Through a case study of 11,010 refactoring and non-refactoring reviews spread across OpenStack open-source projects, we find that refactoring-related code reviews take significantly longer to be resolved in terms of code review efforts. Moreover, upon performing a thematic analysis on a significant sample of the refactoring code review discussions, we built a comprehensive taxonomy consisting of 28 refactoring review criteria. We envision our findings reaffirming the necessity of developing accurate and efficient tools and techniques that can assist developers in the review process in the presence of refactorings.
\end{abstract}





\maketitle

\section{Introduction}
\label{Section:Introduction}


Refactoring is an essential practice to preserve code quality from degradation, as software evolves. Refactoring has grown from the act of cleaning the code, to play a critical role in modern software development. Therefore, refactoring has been attracting various researchers, with over three thousand research papers, according to recent surveys \cite{abid202030}. Another key practice in maintaining software quality is code review \cite{bacchelli2013expectations}. 
 It has become another important to reduce technical debt, and to detect potential coding errors \cite{bacchelli2013expectations,sadowski2018modern,kashiwa2022empirical}. Code review represents the manual inspection of any newly performed changes to the code, for the purpose of verifying integrity, compliance with standards, and error-freedom \cite{mcintosh2016empirical}. Modern code review is a lightweight tool-based process that heavily relies on discussions between commit authors 
  and reviewers to merge/abandon a given code change \cite{yang2016mining}.

Similarly to any other code change, refactoring changes has to also be reviewed before being merged.
  That is, if not applied well, refactoring changes can have their side effects such as hindering software quality \cite{hamdi2021empirical,alomar2019impact,hamdi2021longitudinal,peruma2020exploratory} and inducing bugs \cite{bavota2012does,di2020relationship} 
 making refactoring changes more challenging to review. However, little is known about how reviewers \textit{examine} refactoring related code changes, especially when it is intended to serve the same \textit{purpose} of improving software quality. According to recent industrial case study, AlOmar \etal \cite{alomar2021icse} has found that reviewing  refactoring related code changes takes a significantly longer time, in comparison with other code changes, demonstrating the need of refactoring review \textit{culture}. 
  Yet, little is known about what criteria reviewers consider when they review refactoring. Most of refactoring studies focus on its automation by recommending refactoring opportunities in the source code \cite{tsantalis2008jdeodorant,mkaouer2015many,ouni2016multi}, or mining performed refactorings in change histories of software repositories \cite{tsantalis2018accurate}. 
  Moreover, the research on code reviews has been focused on automating it by recommending the most appropriate reviewer for a given code change \cite{bacchelli2013expectations}. However, despite the the wide adoption of refactoring in practice, its review process is largely unexplored. 

The goal of this paper is to understand what developers care about when they review code, \textit{i.e.,}  what are the main criteria developers rely on to develop a decision about accepting or rejecting a submitted refactored code, and what makes this process challenging. This paper seeks to develop a taxonomy of all refactoring \textit{contexts}, where reviewers are raising concerns about refactoring. We drive our study using the following research questions:

\noindent\textbf{RQ1.} \textit{\RQone}

\noindent\textbf{RQ2.} \textit{\RQtwo} 

To answer these research questions, we first extracted set of 5,505 refactoring-related code reviews, from OpenStack ecosystem. 
 Then, we compared this set of refactoring-related code reviews, with another set of code reviews, in terms of number of reviewers, number of review comments, number on inline comments, number of revision, number of changed files, review duration, discussion and description length, and code churn. 
  Our empirical investigation indicates that refactoring-related code reviews
take significantly longer to be resolved and typically triggers more discussions between developers and reviewers to reach a consensus. To understand the key characteristics of reviewing refactored code, we perform a thematic analysis on a significant sample of these reviews. This process resulted in a hierarchical taxonomy composed of 6 categories, and 28 sub-categories. We also externally validated our taxonomy using a survey of 11 questions related to our categories' correctness and representativeness. We also conducted a follow-up interview to further discuss the survey outcomes.

We provide our comprehensive experiments package \cite{ReplicationPackage} to further replicate and extend our study. The package contains the raw data, analyzed data, statistical test results, survey questions, interview transcription, and custom-built scripts used in our research.

The remainder of this paper is organized as follows: Section \ref{Section:RelatedWork} reviews the existing studies related to refactoring awareness and code review. Section \ref{Section:Methodology} outlines our empirical setup in terms of data collection, analysis and research question. Section \ref{Section:Result} discusses our findings, while the research implication is discussed in Section \ref{Section:Implications}. Section \ref{Section:Threats} captures any threats to the validity of our work, before concluding with Section \ref{Section:Conclusion}.
\section{Related Work}
\label{Section:RelatedWork}

\begin{table*}
  \centering
	 \caption{Related work in refactoring-related code review.}
	 \label{Table:Related_Work_in_Refatoring_Code_Review}
\begin{adjustbox}{width=1.0\textwidth,center}
\begin{tabular}{lclllll}\hline
\toprule
\bfseries Study & \bfseries Year  &  \bfseries Research Type & \bfseries Research Method & \bfseries Purpose & \bfseries Evaluation Technique  \\
\midrule
Ge \etal  \cite{ge2014towards,ge2017refactoring} & 2014,2017 & Tool-based code review & Case \& formative study & Detect refactoring from non-refactoring changes & 2 OSS \& 35 developers  \\ 
Alves \etal  \cite{alves2014refdistiller,alves2017refactoring} & 2014,2018 & Tool-based code review & User study & Inspect manual refactoring edits & 3 OSS \& 15 developers \\ 
Morales \etal  \cite{morales2015code} & 2015 & Empirical study & Case study  & Understand the impact of code review on quality &  3 OSS \\ 
Coelho \etal  \cite{coelho2019refactoring} & 2019 & Literature review & Systematic mapping study & Present refactoring solutions to MCR & N/A \\
Pascarella \etal  \cite{DBLP:journals/corr/abs-1912-10098} & 2019 & Empirical study & Case study & Study the effect of code review on code smells &  7 OSS  \\ 
Paix{\~a}o \etal  \cite{paixao2020behind} & 2020 & Empirical study & Mining-based & Study refactoring in MCR  & 6 OSS\\ 
Pantiuchina \etal \cite{pantiuchina2020developers} & 2020 &  Empirical study & Mining-based & Study refactoring in pull requests  & 150 OSS \\
Uch{\^o}a \etal \cite{uchoa2020does} & 2020 & Empirical study & Mining-based & Study MCR \& design degradation & 7 OSS\\ 
Uch{\^o}a \etal \cite{uchoa2021predicting} & 2021 & Empirical study & Mining-based & Predict design impactful changes in MCR & 7 OSS \\ 
AlOmar \etal \cite{alomar2021icse} & 2021 & Empirical \& industrial case study  & Case study \& survey & Understand refactoring challenges in MCR & 24 developers \\
Brito \& Valente \cite{brito2020refactoring} & 2021 &  Tool-based code review & User study & Detect refactoring during code review & 8 developers\\
Kurbatova \etal \cite{kurbatova2021refactorinsight} & 2021 & Tool-based code review & Mining-based &  Enhance IDE representation
of changes & 4 OSS \\
Coelho \etal \cite{coelho2021empirical} & 2021 &  Empirical study & Mining-based & Study refactoring-inducing pull requests  &  350 OSS \\
\textbf{This work} & \textbf{2022} & \textbf{Empirical \& case study} & \textbf{Mining-based} & \textbf{Understand refactoring practices in MCR} & \textbf{2,225 OSS} \\

\bottomrule
\end{tabular}
\end{adjustbox}
\vspace{-.3cm}
\end{table*}

Research on code review has been of importance to practitioners and researchers. A considerable effort has been spent by the research community in studying traditional and modern code review practices and challenges. This literature has been includes case studies  (\eg,  \cite{ge2014towards,mcintosh2014impact,ge2017refactoring,morales2015code,sadowski2018modern,rigby2013convergent,alomar2021icse}), user studies (\eg, \cite{barnett2015helping,tao2015partitioning,zhang2015interactive,alves2017refactoring,peruma2022refactor}), surveys (\eg,  \cite{tao2012software,bacchelli2013expectations,macleod2017code,alomar2021icse}), and empirical experiments (\eg,  \cite{mcintosh2014impact,morales2015code,tao2015partitioning,guo2017interactively,peruma2019contextualizing}). However, most of the above studies focus on studying and improving the effectiveness of modern code review in general, as opposed to our work that focuses on understanding developers' perception of code review involving refactoring. In this section, we are only interested in research related to refactoring-aware code review. We summarize these approaches in Table~\ref{Table:Related_Work_in_Refatoring_Code_Review}.

In a study performed at Microsoft, Bacchelli and Bird \cite{bacchelli2013expectations} observed, and surveyed developers to understand the challenges faced during code review. They pointed out purposes for code review (\eg, improving team awareness and transferring knowledge among teams) along with the actual outcomes (\eg, creating
awareness and gaining code understanding). In a similar context, MacLeod \etal  \cite{macleod2017code} interviewed several teams at Microsoft and conducted a survey to investigate the human and social factors that influence developers' experiences with code review. Both studies found the following general code reviewing challenges: (1) finding defects, (2) improving the code, and (3) increasing knowledge transfer. Ge \etal  \cite{ge2014towards,ge2017refactoring} developed a refactoring-aware code review tool, called ReviewFactor, that automatically detects refactoring edits and separates refactoring from non-refactoring changes with the focus on five refactoring types. The tool was intended to support developers' review process by distinguishing between refactoring and non-refactoring changes, but it does not provide any insights on the quality of the performed refactoring. Inspired by the work of \cite{ge2014towards,ge2017refactoring}, Alves \etal  \cite{alves2014refdistiller,alves2017refactoring} proposed a static analysis tool, called RefDistiller, that helps developers inspect manual refactoring edits. The tool compares two program versions to detect refactoring anomalies' type and location. It supports six refactoring operations, detects incomplete refactorings, and provides inspection for manual refactorings. 


Coelho \etal  \cite{coelho2019refactoring} performed a systematic literature mapping study on refactoring tools to support modern code review. They raised the need for more tools to explain composite refactorings.  They also reported the need for more surveys to assess the existing refactoring tools for modern code review in both open source and industrial projects. Pascarella \etal  \cite{pascarella2018information} investigated the effect of code review on bad programming practices (\ie, code smells). Their approach mainly focused on comparing code smells at file-level before and after the code review process. Additionally, they manually investigated whether the severity of code smells was reduced in a code review or not. Their results show, in 95\% of the cases, the severity of code smells does not decrease with a review. The reduction of code smells in remaining few cases was impacted by code insertion and refactoring-related changes. 

Paix{\~a}o \etal  \cite{paixao2020behind} explored if developers’ intents influence the evolution of refactorings during the review of a code change by mining 1,780 reviewed code changes from 6 open-source systems. Their main findings show that refactorings are most often used in code reviews that implement new features, accounting for 63\% of the code changes we studied. Only in 31\% of the code reviews that employed refactorings the developers had the explicit intent of refactoring. Uch{\^o}a \etal \cite{uchoa2020does} reported the multi-project retrospective study that characterizes how the process of design degradation evolves within each review and across multiple reviews. The authors utilized software metrics to observe the influence of certain code review practices on combating design degradation. The authors found that the majority of code reviews had little to no design degradation impact in the analyzed projects. Additionally, the practices of long discussions and high proportion of review disagreement in code reviews were found to increase design degradation. In their study on predicting design impactful changes in modern code review with technical and/or social aspects, Uch{\^o}a \etal \cite{uchoa2021predicting} analyzed reviewed code changes from seven open source projects. By evaluating six machine learning algorithms, the authors found that technical features results in more precise predictions and the use of social features alone also leads to accurate predictions.

A couple of studies considered pull requests as a main source of the study code review process. Pantiuchina \etal \cite{pantiuchina2020developers} presented a mining-based study to investigate why developers are performing refactoring in the history of 150 open source systems. Particularly, they analyzed 551 pull requests implemented refactoring operations and reported a refactoring taxonomy that generalizes the ones existing in the literature. Coelho \etal \cite{coelho2021empirical} performed a quantitative and qualitative study exploring code reviewing-related aspects intending to characterize refactoring-inducing pull requests. Their main finding show that refactoring-inducing pull requests take significantly more
time to merge than non-refactoring-inducing pull requests.

AlOmar \etal \cite{alomar2021icse} conducted a case study in an industrial setting to explore refactoring practices in the context of modern code review from the following five dimensions: (1) developers motivations to refactor their code, (2) how developers document their refactoring for code review, (3) the challenges faced by reviewers when reviewing refactoring changes, (4) the mechanisms used by reviewers to ensure the correctness after refactoring, and (5) developers and reviewers assessment of refactoring impact on the source code's quality. Their findings show that refactoring code reviews take longer to be completed than the non-refactoring code reviews. 
Brito \& Valente \cite{brito2020refactoring} introduced RAID, a refactoring-aware and intelligent diff tool to alleviate the cognitive effort associated with
code reviews. The tool relied on RefDiff \cite{silva2020refdiff} and is fully integrated with the state-of-the-art practice of continuous integration pipelines (GitHub Actions) and browsers (Google Chrome). The authors evaluated the tool with eight professional developers and found that RAID indeed reduced the cognitive effort required for detecting and reviewing refactorings. In another study, Kurbatova \etal \cite{kurbatova2021refactorinsight} presented RefactorInsight, a plugin for IntelliJ IDEA that integrates information about refactorings in diffs in the IDE, auto folds refactorings in code diffs in Java and Kotlin, and shows hints with their short descriptions.

To summarize, the study of open source projects that use either the Gerrit tools or GitHub pull
requests has been extensively studied (\eg, \cite{rigby2013convergent,thongtanunam2015should,zhang2018multiple, pantiuchina2020developers}). Since notable open source organizations such as Eclipse and OpenStack adopted Gerrit as their code review management tool, we chose to analyze refactoring practice in modern code review from projects that adopted Gerrit as their code review tool. 
Although there are recent studies that explored the motivation behind refactoring in pull requests \cite{pantiuchina2020developers,coelho2021empirical}, to the best of our knowledge, no prior studies have manually extracted all the criteria developers are facing when submitting their refactored code for review. To gain more in-depth understanding of factors mostly associated with refactoring review discussion and to advance the understanding of refactoring-aware code review, in this paper, we performed an empirical study on a rapidly evolving open source project, with large number of files with 100\% review coverage. This study complements the existing efforts that are done in an industrial environment \cite{alomar2021icse} and an open source systems \cite{pantiuchina2020developers,coelho2021empirical} using GitHub pull-based development.

\section{Study Design}
\label{Section:Methodology}

The main goal of our study is to understand refactoring practice in the context of modern code review to characterize the criteria that influence the decision making when reviewing refactoring changes. Thus, we aim at answering the following research questions:

\begin{itemize}
\item \textbf{RQ1.} \textit{\RQone}

\item \textbf{RQ2.} \textit{\RQtwo}
\end{itemize}

\begin{figure*}[t]
 	\centering
 	\includegraphics[width=1.0\textwidth]{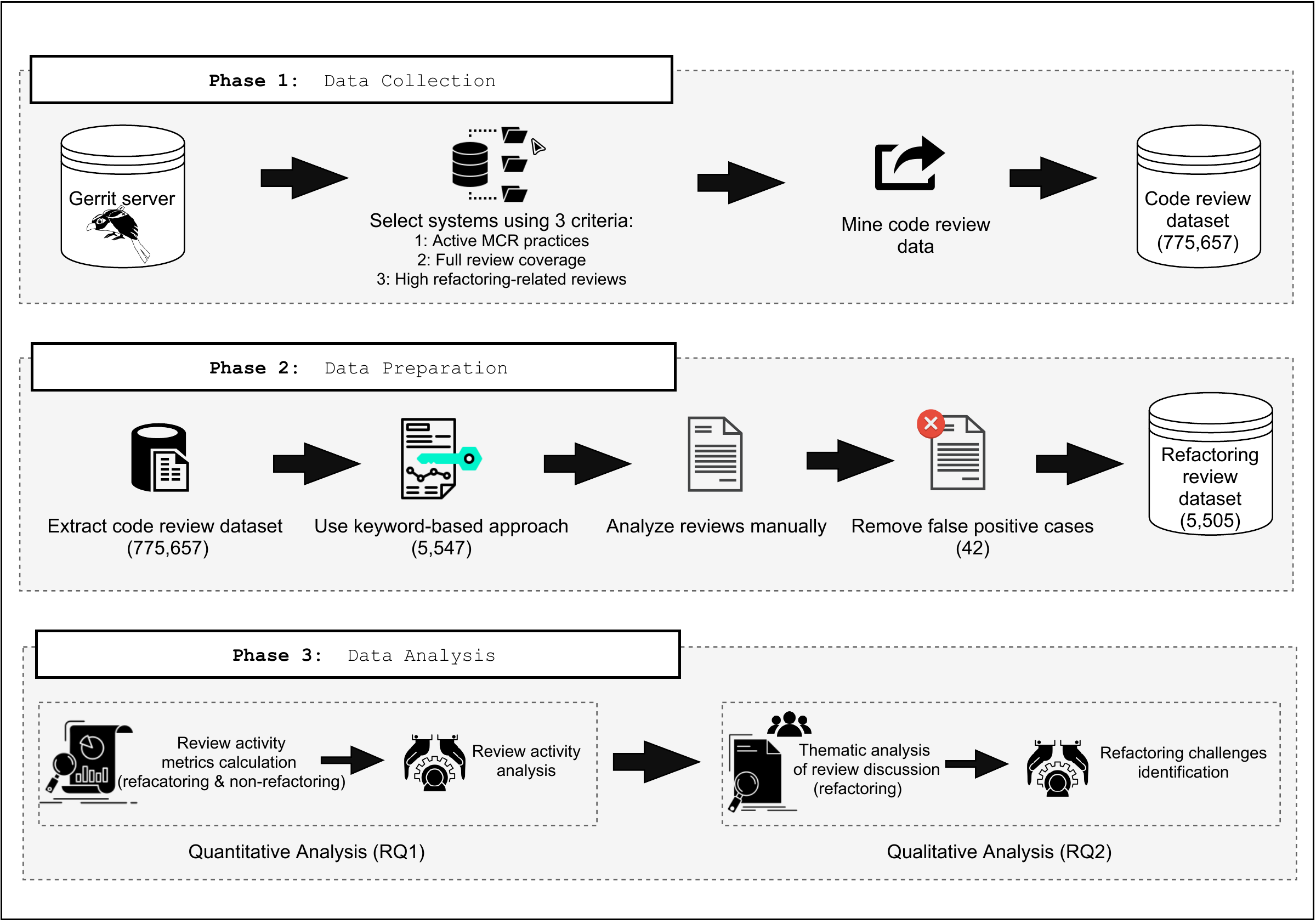}
 	\caption{Overview of our experiment design.}
 	\label{fig:approach}
\end{figure*}

According to the guidelines reported by Runeson and H{\"o}st \cite{runeson2009guidelines}, we designed an empirical study that consists of three steps, as depicted in Figure \ref{fig:approach}. 
  Since our research questions are both quantitative and qualitative, we used tools/scripts along with manual activities to investigate our data. 
Furthermore, the dataset utilized in this study is available on our project website \cite{ReplicationPackage} for extension and replication purposes.

\begin{figure}[]
 	\centering
 	\includegraphics[width=1.0\columnwidth]{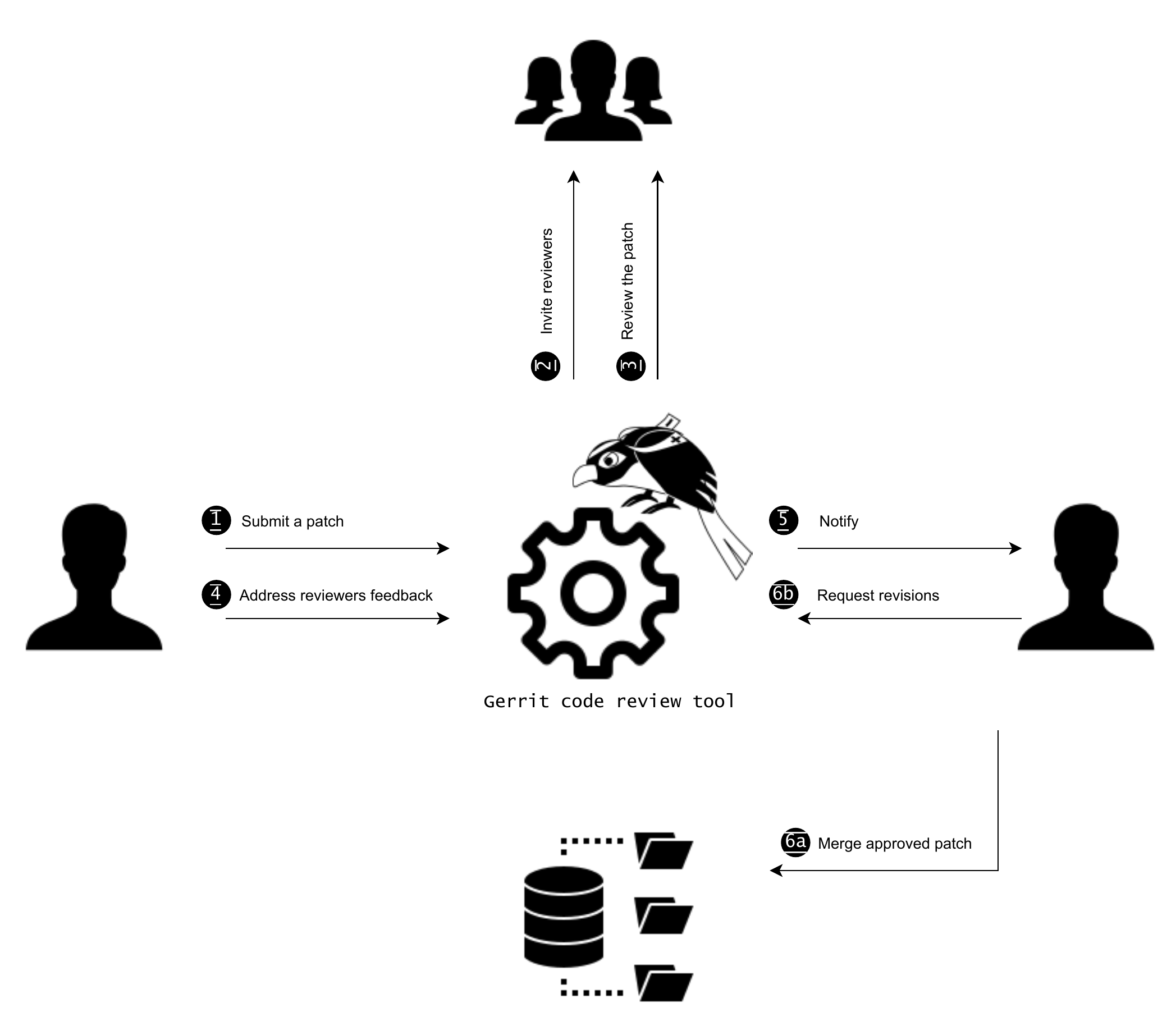}
 	\caption{Gerrit-based code review process overview.} 
 	\label{fig:review}
\vspace{-.4cm}
\end{figure}

\noindent \textbf{Gerrit-based code review process.} The code review process of the studied systems is based on Gerrit\footnote{\url{https://www.gerritcodereview.com/}}, collaborative code review framework allowing developers to directly tag submitted code changes and request its assignment to a reviewer. Generally, a code change author opens a code review request containing a title, a detailed description of the code change being submitted, written in natural language, along with the current code changes annotated. Once the review request is submitted, it appears in the requests backlog, open for reviewers to choose. Once reviewers are assigned to the review request,
they inspect the proposed changes and comment on the review request's thread, to start a discussion with the author. This way, the authors and reviewers can discuss the submitted changes, and reviewers can request revisions to the code being reviewed. Following up discussions and revisions, a review decision is made to either accept or decline, and so the proposed code changes are either \say{\textit{Merged}} to production or \say{\textit{Abandoned}}. A diagram, modeling a simplified bird's view of the Gerrit-based code review process, is shown in Figure \ref{fig:review}. 

\subsection{Data Collection}
\subsubsection{Studied Systems}
To select the subject systems, we identified three important criteria:

\noindent \textbf{Criterion \#1: Active MCR practices.} Our goal is to study a system that actively examines code changes through a code review tool. Therefore, we focus on systems where a number of reviews are performed using a code review tool \textcolor{black}{(\ie, systems which have review procedures in place)}, similar to \cite{thongtanunam2015investigating,thongtanunam2016revisiting,morales2015code}. 

\noindent \textbf{Criterion \#2: Full review coverage.}
Since we investigate the practice of refactoring-related code reviews, we focus on  systems that have many files with 100\% review coverage \textcolor{black}{(\ie, files where every change made to them is reviewed before they are merged into the repositories)}, similar to the studies that explored code review practices in defective files \cite{thongtanunam2015investigating,thongtanunam2016revisiting,mcintosh2014impact}. 

\noindent \textbf{Criterion \#3: High number of refactoring-related reviews.} Since we want to study refactoring practices in MCR, we need to ensure that the subject systems have sufficient refactoring-related instances to help us perform our statistical analysis. \textcolor{black}{So, we selected the project with the highest number of refactoring reviews.}


To satisfy criterion 1, we started by considering five systems (\ie,  OpenStack, \footnote{https://review.opendev.org/} Qt ,\footnote{ https://codereview.qt-project.org/} LibreOffice, \footnote{https://gerrit.libreoffice.org/} VTK, \footnote{http://vtk.org/} ITK \footnote{http://itk.org/}) 
that use Gerrit code review tool and have been widely studied in previous research
in MCR, \eg, \cite{thongtanunam2020review,ouni2016search,fan2018early,chouchen2021whoreview}. We then discarded VTK and ITK since Thongtanunam \etal \cite{thongtanunam2016revisiting} reported that the linkage rate of code changes to the reviews for VTK is too low and ITK does not satisfy criterion 2.  As for criterion 3, after mining the code review data, we found that OpenStack has a higher number of refactoring-related code review instances than Qt and LibreOffice. Due to the human-intensive nature of carefully studying and analyzing refactoring practice in MCR, we opt for performing an in-depth study on a single system. With the above-mentioned criteria in mind, we select OpenStack, 
 an open-source software for cloud infrastructure service that is developed by many well-known companies, \eg, IBM, VMware, and NEC.

\subsubsection{Mining code review data}
We mined code review data using the \texttt{RESTful API}\footnote{https://gerrit-review.googlesource.com/Documentation/rest-apichanges.html} provided by Gerrit, which returns the results in a \texttt{JSON} format. We used a script to automatically mine the review data and store them in \texttt{SQLite} database. All collected reviews are closed (\ie,  having a status of either \say{\textit{Merged}} or \say{\textit{Abandoned}}). 
In total, we mined 775,657 code changes between December 2012 and April 2021 from OpenStack projects. An overview of the project's statistics is provided in Table~\ref{Table:DATA_Overview}.
\begin{table}[h!]
\begin{center}
\caption{Overview of the OpenStack studied system.} 
\label{Table:DATA_Overview}
\begin{adjustbox}{width=0.5\textwidth,center}
\begin{tabular}{lllll}\hline
\toprule
\bfseries Item & \bfseries Count \\
\midrule
Review period & 12.16.2012 to 04.27.2021\\
Number of projects & 2,225 \\
Version & Folsom to Stein \\
Line of code & 31,680,274 \\
No. of commits & 4,137,446 \\
No. of code changes &  775,657 \\
No. of developers & 15,432  \\
No. of files &  705,982  \\
Reviews with keyword `\textit{refactor*}' in title or description & 10,440 \\
Reviews with keyword `\textit{refactor*}' in title and description & 5,547  \\
False positive reviews & 42 \\
Refactoring review dataset & 5,505 \\
\textcolor{black}{Final refactoring and non-refactoring reviews dataset} & 11,010 \\
\bottomrule
\end{tabular}
\end{adjustbox}
\end{center}
\end{table}

\subsection{Data Preparation}



To extract the set of refactoring-related code reviews, we follow a two-step procedure: (1) automatic filtering, and (2) manual filtering. 

\textbf{(1) Automatic Filtering.} In the first step, we utilize a keyword-based mechanism to filter out all entries 
that do not contain the keyword \textit{refactor*} in both the title and description of the submitted code change. Specifically, we start by searching for the term `\textit{refactor*}' in the title or description (we use * to capture extensions like refactors, refactoring etc.). The keyword-based approach has been widely used in prior studies related to identify refactoring changes or defect-fixing or defect-inducing changes \cite{kim2008classifying,kamei2012large,mockus2000identifying,hassan2008automated,thongtanunam2016revisiting,mcintosh2016empirical,pantiuchina2020developers,Ratzinger:2008:RRS:1370750.1370759,coelho2021empirical,alomar2019can,alomar2021ESWA,stroggylos2007refactoring, alomar2019impact,tang2021empirical,alomar2021documentation,alomar2021toward,alomar2020developers,alomar2021refactoringreuse}, as it allows the pruning of the search space to only consider code changes whose documentation matches a specific intention. The choice of `\textit{refactor}', besides being used by various related studies, is intuitively the first term to identify ideal refactoring-related code changes \cite{kim2008classifying,alomar2021ESWA,alomar2021toward,alomar2021behind,alomar2021documentation}. Also, the choice of enforcing the existence of the keyword in both code change's 
 title or description was piloted by a first trial of only applying the filter to description. This initial filtering gives us a total of 10,440 review instances. After performing a manual inspection of a subset of these reviews, we realized that there exist several false positive cases. To reduce them, we only kept code changes 
having the term `\textit{refactor*}' in both the title and description. The keyword-based filtering resulted in only selecting 5,547 code changes 
and their corresponding reviews. We notice that the ratio of these reviews is very small in comparison with the total number of the mined reviews, \ie, 775,657. However, these observations aligned with previous studies \cite{murphy2012we,szoke2014bulk} as developers typically do not provide details when documenting their refactorings. Yet, despite this strict filtering, this approach is still prone to high false positiveness, and therefore, the second step of manual analysis is needed.

\textbf{(2) Manual Filtering.} To ensure the correctness of data, we manually inspected and read through all these refactoring reviews to remove false positives. An example of a discarded review is: \say{\textit{Revert "Refactor create\_pool." and "Add request\_access\_to\_group method"}}  
\cite{Example-1}. We discarded this code review because the refactoring action is undone by developers. This step resulted in only considering 5,505 reviews. Our goal is to have a \textit{gold set} of reviews in which the developers explicitly reported the refactoring activity. This \textit{gold set} will serve to check later criteria that are mostly associated with refactoring review discussion. 


\subsection{Data Analysis}
To address our research questions, a structured mixed-method study was designed to combine elements of both quantitative and qualitative research.

\subsubsection{Quantitative data analysis.} We leverage the data collected to compare refactoring and non-refactoring reviews using review efforts, \ie, code review metrics.
 As we calculate the metrics of refactoring and non-refactoring code reviews, we want to distinguish, for each metric, whether the variation is statistically significant. 
 We first test for normality using Shapiro-Wilk normality test \cite{taeger2014statistical} and we observe that the distribution of code review activity metrics does not follow a normal distribution. Therefore, we use the Mann-Whitney U test \cite{conover1998practical}, a non-parametric test, to compare between the two groups, since these groups are independent of one another. The null hypothesis is defined by no variation in the metric values of refactoring and non-refactoring code reviews. Thus, the alternative hypothesis indicates that there is a variation in the metrics values. Additionally, the variation between values of both sets is considered significant if its associated \textit{p}-value is less than 0.05. Furthermore, we use the Cliff's Delta ($\delta$) \cite{cliff1993dominance}, a non-parametric effect size measure, to estimate the magnitude of the differences between refactoring and non-refactoring reviews. As for its interpretation, we follow the guidelines reported by Romano \etal \cite{romano2006appropriate}:
 
 \begin{itemize}
\item Negligible for $\mid \delta \mid< 0.147$
\item Small for $0.147 \leq \mid \delta \mid < 0.33$
\item Medium for $0.33 \leq \mid \delta \mid < 0.474$
\item Large for $\mid \delta \mid \geq 0.474$
\end{itemize}
 
 To measure the extent of the relationship between these metrics, we conducted a Spearman rank correlation test (a non-parametric measure) \cite{wissler1905spearman}. We chose a rank correlation because this type of correlation is resilient to data that is not normally distributed. 

\subsubsection{Qualitative data analysis.} To answer RQ2, two of the authors perform the analysis of the data. One of the author manually inspects refactoring review discussions \textcolor{black}{by considering both the general comments and the inline comments}, and the other author reviews the taxonomy. As the complete list of refactoring review data is too large to be manually examined, we select a statistically significant sample for our analysis by considering the reviews with higher review duration, and we annotated 384 reviews. This quantity roughly equates to a sample size with a confidence level of 95\% and a confidence interval of 5. The manual analysis process took approximately 40 days in total. Next, we describe the methodology for building and refining the taxonomy, followed by the validation method. 

\vspace{.1cm}
{\textbf{Taxonomy Building and Refinement.}} When analyzing the review discussions, we adopted a thematic analysis approach based on guidelines provided by Cruzes \etal \cite{cruzes2011recommended}. Thematic analysis is one of the most used methods in Software Engineering literature (\eg, \cite{Silva:2016:WWR:2950290.2950305}), which is a technique for identifying and recording patterns (or \say{themes}) within a collection of descriptive labels, which we call \say{codes}. For each refactoring review, we proceeded with the analysis using the following steps: 
i) Initial reading of the review discussions; ii) Generating initial codes (\ie, labels) for each review; iii) Translating codes into themes, sub-themes, and higher-order themes; iv) Reviewing the themes to find opportunities for merging; v) Defining and naming the final themes, and creating a model of higher-order themes and their underlying evidence.

The above-mentioned steps were performed independently by two authors. \textcolor{black}{One author performed the labeling  of review discussions independently from the other author who was responsible for reviewing the currently drafted taxonomy. By the end of each iteration, the authors met and refined the taxonomy}. At the time of the study, \textcolor{black}{one of the
author} had 4 years of research experience on refactoring, while \textcolor{black}{the other author} had 9 years of research experience on refactoring.

It is important to note that the approach is not a single step process. As the codes 
 were analyzed, some of the first cycle codes 
were subsumed by other codes, relabeled, or dropped all together. As the two authors progressed in the translation to themes, there was some rearrangement, refinement, and reclassification of data into different or new codes. For example, we aggregated, into \say{\textit{Refactoring}}, the preliminary categories \say{\textit{incorrect refactoring}}, \say{\textit{behavior preservation violation}}, \say{\textit{separation of other changes from refactoring}}, \say{\textit{interleaving other changes with refactoring}}, and \say{\textit{domain constraint}} that were discussed by different reviewers. We used the thematic analysis technique to address RQ2.  

\vspace{.1cm}
{\textbf{\textcolor{black}{Taxonomy Validation.}}} \textcolor{black}{In addition to the iterative process of building the taxonomy, we need to also externally validate it from a practitioner's point of view \cite{pascarella2018self,dougan2022towards}. The aim of this validation is to investigate whether it reflects actual MCR practices. To do so, we validated the taxonomy with a senior developer, with 8 years of industrial and refactoring experience, and with 6 years of experience in code review. The survey contained 11 questions related to the correctness and representativeness of our taxonomy. We also conducted a follow-up interview to further discuss the survey outcomes. The interview took an hour and was recorded for further analysis. 
The interview summary is available in the replication package.}

\section{Results and Discussion}
\label{Section:Result}
\begin{table*}
\centering
\caption{Statistics of code review activity efforts.} 
\label{Table:RQ1_results}
\begin{adjustbox}{width=1.0\textwidth,center}
\begin{tabular}{@{}lllllll|llllll|ll@{}}
\toprule
\multicolumn{1}{c}{\multirow{3}{*}{\textbf{Metrics}}} & \multicolumn{6}{c|}{\textit{\textbf{Refactoring code review}}} & \multicolumn{6}{c|}{\textit{\textbf{Non-refactoring code review}}} & \multicolumn{2}{c}{\textit{\textbf{Statistical difference}}}  \\ \cmidrule(l){2-15} 
\multicolumn{1}{c}{} & \multicolumn{1}{c}{\textbf{Min}} & \multicolumn{1}{c}{\textbf{Q1}} & \multicolumn{1}{c}{\textbf{Median}} & \multicolumn{1}{c}{\textbf{Mean}} & \multicolumn{1}{c}{\textbf{Q3}} & \multicolumn{1}{c|}{\textbf{Max}} & \multicolumn{1}{c}{\textbf{Min}} & \multicolumn{1}{c}{\textbf{Q1}} & \multicolumn{1}{c}{\textbf{Median}} & \multicolumn{1}{c}{\textbf{Mean}} & \multicolumn{1}{c}{\textbf{Q3}} & \multicolumn{1}{l|}{\textbf{Max}} &
\multicolumn{1}{c}{\textbf{\textit{p}-value}} & \multicolumn{1}{l}{\textbf{Cliff's delta ($\delta$)}}\\ 
\midrule
Number of reviewers & 0 & 2 & 4 &  5.65 & 6  & 12 & 0 &2 &3  &4.45  &5  & 9 &  \textbf{6.750458e-43} & small (0.15)\\
Number of review comments & 0 & 5 & 9 & 20.87 & 19 & 40 & 0 & 3 & 6 & 13.05  & 12 & 25 & \textbf{2.730193e-82} & small (0.22) \\
Number of inline comments & 0 & 0 & 0 & 6.61 & 5 & 12 & 0 & 0 & 0 & 3.26 & 1 & 2 & \textbf{2.066649e-98} & medium (0.33) \\
Number of revisions & 1 & 1 & 3 & 4.79 & 6 & 13 & 1 & 1 & 2 & 3.19 & 3 & 6 & \textbf{1.252695e-126} & small (0.3)\\
Number of changed files & 0 & 1 & 3 & 5.98 & 6 & 13 & 0 & 1 & 1 & 3.69 & 3 & 6 &  \textbf{6.006899e-190} & medium (0.33)  \\
Review duration (seconds) & 0 & 37  & 170.68 & 928.21 & 606.22 & 1458.68 & 0 & 12.85 & 75.98 & 738.93 & 351.03 & 852.99 & \textbf{2.572584e-66} & small (0.23) \\
Length of discussion (characters)   & 0 & 168 & 451 & 3450.29 & 1563 & 3653 & 0 & 110 & 296 & 1962.39 & 1044 & 2436 &  \textbf{6.966752e-42} & small (0.15)\\
Length of description (characters) & 62 & 171 & 264 & 327.61 & 388 & 711 & 56 & 124 & 218 & 276.97 & 354 & 699 & \textbf{4.071541e-41} & small (0.13) \\
Code churn  & -1 & 40 & 114 & 367.63 & 309 & 711 & 0 & 3 & 11 & 201.58 & 53 & 128 & \textbf{0.000000e+00} &  large (0.64) \\
\bottomrule
\end{tabular}
\end{adjustbox}
\end{table*}
\subsection{\RQone} 



\noindent\textbf{Approach.} To address RQ1, we intend to compare \textit{refactoring reviews} with \textit{non-refactoring reviews}, to see whether there are any differences in terms of code review efforts or metrics listed in Table \ref{Table:RQ1_results}. 
Since our refactoring set contains 5,505 reviews, we need to sample 5,505 non-refactoring reviews from the remaining ones in the review framework. To ensure the representativeness of the sample \cite{clarkson1989applications}, we use  stratified random sampling by choosing reviews from the rest of reviews.

\noindent\textbf{Results.} By looking at the statistical summary in Table \ref{Table:RQ1_results}, we found that reviewing refactoring changes significantly differ (\ie, more reviewers ($\mu$ = 5.56), more review comments ($\mu$ = 20.87), more inline comments ($\mu$ = 6.61), more revisions ($\mu$ = 4.79), more file changes ($\mu$ = 5.98), lengthier review time ($\mu$ = 928.21), discussions and descriptions ($\mu$ = 3450.29, $\mu$ = 327.61, respectively), and more added and deleted lines between revisions ($\mu$ = 367.63) from reviewing non-refactoring changes. As shown in Table \ref{Table:RQ1_results}, we performed a non-parametric Mann-Whitney U test and we obtained a statistically significant \textit{p}-value when the values of these two groups were compared (\textit{p}-value $< 0.05$ for all review efforts), and accompanied with a small, medium, or large effect size depending on the review effort/metric. 




We speculate that reviewing refactoring triggers longer discussions between the code change authors and the reviewers as we notice that several refactoring-related actions are being extensively discussed before reaching an agreement. While previous studies have found a similar pattern in GitHub's pull requests in open-source systems \cite{coelho2021empirical} and using code review tools in industry \cite{alomar2021icse}, there is not a study that looked at what are the main reasons for refactoring-related discussions to take significantly longer effort to be reviewed. Therefore, the findings of RQ1 has motivated us to manually analyze these reviews and extract the main criteria related to reviewing refactored code (RQ2). 

Further, we observe that refactoring related code reviews impact larger code churn and more changes across files than non-refactoring code changes. These results are expected and are in agreement with prior works \cite{coelho2021empirical,hegedHus2018empirical,paixao2019impact}, which found that refactored code has higher size-related metrics and larger changes promote refactorings. We also notice that the number of developers who participated in the refactoring code review process is also higher due to the high number of added, modified, or deleted lines between revisions. In contrast to a previous finding \cite{coelho2021empirical}, however, no evidence of the correlation between the number of reviewers and refactoring was detected. 

Moreover, our correlation analysis is measured using the Spearman rank correlation test reveals that the number of review comments, discussion length, and number of revisions are  highly correlated with the review duration. The Spearman rank correlation test yielded a statistically significant (\ie, \textit{p}-value $< 0.05$) correlation coefficient of 0.57, 0.53 and 0.49, respectively. Further, the number of reviewers are highly correlated with the discussion length and number of review comments with \textit{p}-value $< 0.05$ and correlation coefficient of 0.73 and 0.77, equating to a strong correlation. Additionally, we observe that very high number of deleted or added lines are correlated with very high number of file changes (\ie, \textit{p}-value $< 0.05$ and correlation coefficient of 0.58). In contrast, the Spearman’s correlation values detect no significance in the relationship between the number of files and review duration. 

\subsection{What challenges do developers face when reviewing refactoring tasks?}
\noindent\textbf{Approach.} To get a more qualitative sense, we manually inspect the OpenStack ecosystem using a thematic analysis technique \cite{cruzes2011recommended}, to study the challenges that reviewers catch when reviewing refactoring changes, so we understand the main reasons for which refactoring reviews take longer compared to non-refactoring reviews. 

\begin{figure*}[t]
\centering 
\includegraphics[width=\textwidth]{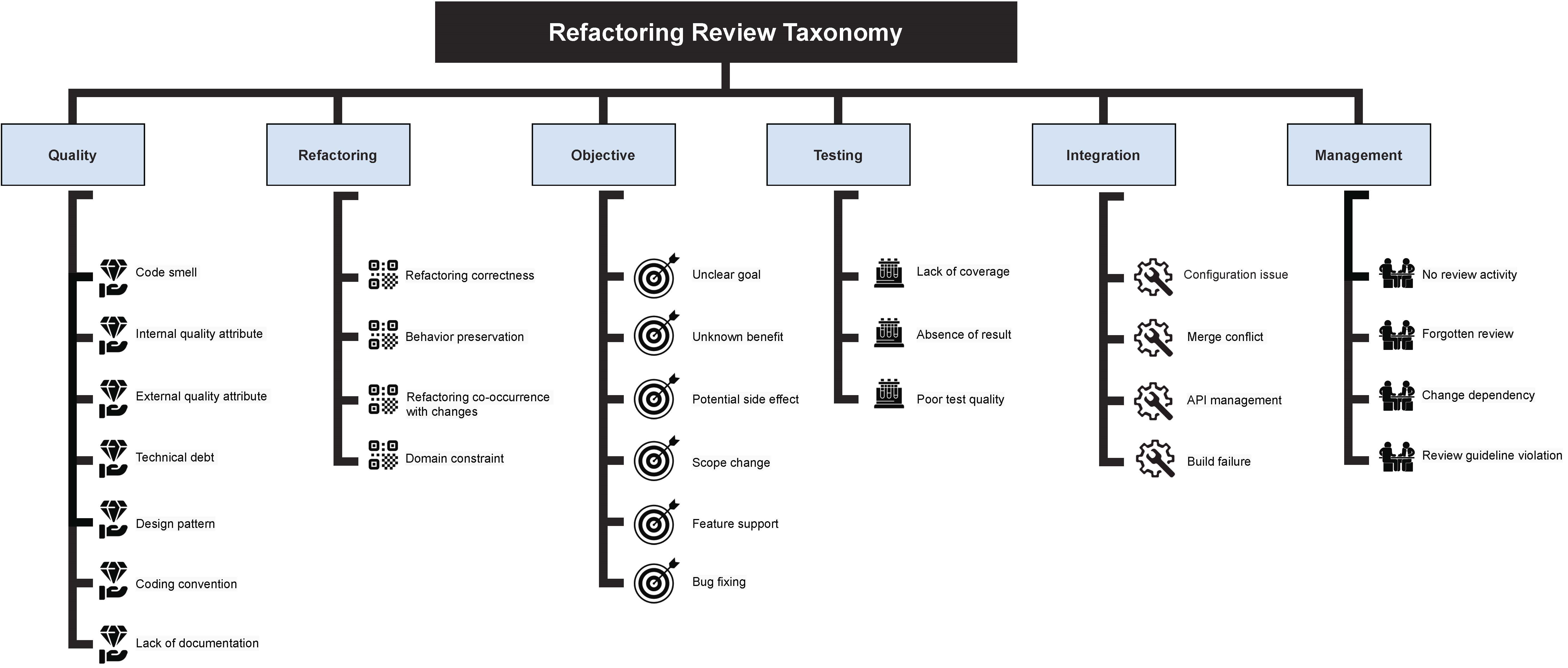}
\caption{Refactoring review criteria in modern code review.}
\label{fig:challenges}
\end{figure*}

\noindent\textbf{Results.} Upon analyzing the review discussions, we create a comprehensive high-level categories of review criteria. Figure \ref{fig:challenges} shows the proposed taxonomy of the criteria related to reviewing refactored code. The taxonomy is composed of two layers: the top layer contains 6 categories that group activities with similar purposes, whereas the lower layer contains 28 subcategories that essentially provide a fine-grained categorization. Due to space constraints, we made the iterative taxonomy available in our replication package \cite{ReplicationPackage}. 
 These refactoring review criteria 
  are centered around six main categories 
  as shown in the figure: (1) quality, (2) refactoring 
  , (3) objective 
  , (4) testing, (5) integration, and (6) management. 
   It is worth noting that our categorization is not mutually exclusive, meaning that a review can be associated with more than one category. An example of each category is provided in Table \ref{Table:example}. 
\textcolor{black}{Further, according to the interview with the senior developer, they conformed that our taxonomy is representative of the reviews cases they have experienced.} 
In the rest of this subsection, we provide more in-depth analysis of these categories.

\noindent\textbf{Category \#1: Quality.} 
The design quality is found to be a vital part of the refactoring review process. According to the review discussions, reviewers enforce the adherence to \textit{coding convention}, optimization of \textit{internal quality attribute}, \textit{external quality attribute}, the avoidance of (\ie, \textit{code smell}, resolution of \textit{technical debt}, correctness of \textit{design pattern} implementations), and \textit{lack of documentation}. For instance, developers recommend appropriate ways to write code, optimize \textit{internal and external quality attributes} as developers may not \textit{draw the full picture} of the software design, which makes their decision adequate locally, but not necessarily at the global level. Moreover, developers only reason on the actual \textit{snapshot} 
 of the current design, and there is no systematic way for them to recognize its evolution by, for instance, accessing previously performed refactorings. This may also narrow their decision making, and they may end up \textit{reverting} some previous refactorings.  Moreover, providing a clear and meaningful explanation of the reasons behind the proposed changes are equally important in review decisions. The requested documentation change includes (but is not limited to) expanding code comments, removing code smells, and reusing existing functionality. 
 Further, developers typically refactor classes and methods that they frequently change. We observe this as various reviews were containing similar recurrent files. 
 So, the more they change the same code elements, the more familiar with the system they become, and so it improves their design decisions. 
\textcolor{black}{While the follow-up interview indicates the importance of each sub-category in real life, it was pointed out by the senior developer 
that \textit{technical debt} 
was not allowed to be submitted since the company has a strict policy preventing developers from submitting near optimal code. 
Additionally, according to the senior developer's experience, reviewers did allow the relaxation of design patterns strict implementations, as long as it is properly justified.} 

\noindent\textbf{Category \#2: Refactoring. 
}  
This category gathers reviews with a focus on evaluating the correctness of the code transformation and checking whether or not the submitted changes lead to a safe and trustworthy refactoring. These reviews discuss  \textit{refactoring correctness}, \textit{behavior preservation}, \textit{refactoring co-occurrence with changes}, and \textit{domain constraint}. Because developers interleave refactorings with other tasks, reviewers highlighted that mixing refactoring with other changes may lead to overshadowing errors, and so the intrusion of bugs. 
\textcolor{black}{Similar concerns were also raised by the senior developer during our interview. In their regular reviews, they would typically recommend the separation of refactoring from other code changes, whenever possible. } 

\noindent\textbf{Category \#3: Objective.} 
In this category, we have gathered cases where reviewers eventually ask to clearly document the \textit{goal}, \textit{benefit}, \textit{side effects}, \textit{scope}, \textit{feature-related}, and \textit{bug fix-related} activities to better understand the rationale of the submitted code changes. This reveals how reviewers keep proposing areas of improvement in developers' documentation practices, pertaining to the perception and the rationale of the change. It appears that the clarity of the documented changes is of paramount importance as reviewers also struggle with identifying the \textit{benefit} and/or \textit{side effects} and spend time understanding them. We realized that the clarity of the explanation of what is being changed and why affects review time and decision. Missing rationale and documentation are also frequently reported as confusing during code review \cite{ebert2021exploratory}. Reviewers are requested to \textit{clarify the scope of the change}. Changing the scope is one of the influential factors for reviewers to make their decisions. Reviewers check the actual appropriateness of the change before it is merged into the code base. Further, we realized that source code is not the only artifact that developers refactor. Other artifacts, such as database elements, are also subject to refactoring. \textcolor{black}{During the follow up interview, the senior developer indicated that \textit{unclear goal} and \textit{unknown benefit} were less likely to be encountered due to the company review guidelines. Also \textit{bug fixing} is not frequent in the expert's current company because of the enforcement of performing minimal refactoring changes when fixing bugs.} 

\noindent\textbf{Category \#4: Testing.} Refactoring is supposed to preserve the behavior of the software. Ideally, using the existing unit tests to verify that the behavior is maintained should be sufficient. However, since refactoring can also be interleaved with other tasks, then there might be a change in the software's behavior, and so, unit tests may not capture such changes if they were not revalidated to reflect the newly introduced functionality. We have seen various reviews discussions raising this concern, especially when developers are unaware of such non behavior preserving changes, and so, deprecated unit tests will not guarantee the refactoring correctness. Based on our analysis of the discussion, reviewers suggested adding unit tests before the refactoring, so that they can be more confident the code has not broken anything. They also recommend adding test cases when the refactored code lowers the test coverage (\eg, extracting new methods). Moreover, when developers submit a review, they can include the results of running the tests as reviewers expect code changes to be accompanied by a corresponding test change. So, to capture these various cases, under the \textit{Testing} category, we have the following sub-categories: \textit{lack of coverage},  \textit{absence of result},  and \textit{poor test quality}. \textcolor{black}{The outcome of the interview shows the existence of this type of review criteria.} 

\noindent\textbf{Category \#5: Integration.} This category gathers reviews highlighting how refactoring has
complicated the merging process, or triggered configuration issues. Several sub-categories raise, 
namely, \textit{configuration issue}, \textit{merge failure}, \textit{API management}, and \textit{build failure}. These categories show that refactoring may complicate the review process. As per our analysis, it appears that the code changes involving refactoring reviews tend to be more problematic for failure or issue, compared to non-refactoring reviews. This observation is partially in lined with previous studies \cite{mahmoudi2019refactorings,dig2007refactoring}. They found that merge conflicts that involve refactoring are more complex than conflicts with no refactoring changes. Moreover, we noticed that API upgrades and migrations typically trigger discussions, mainly when there are API breaking changes. Developers tend to ask about the appropriate migration plans and the necessary changes to preserve the software's behavior during the migration process. Despite the existence of various tools to support the detection of breaking changes, discussions revealed that this process is still manual. \textcolor{black}{Besides agreeing with the existence of these types subcategories, the participant also reported that his current company supports developers with review \textit{bots} to early detect any merge-related issues.} 
\begin{table*}[htbp]
  \centering
	 \caption{A taxonomy of the refactoring review criteria in Modern Code Review.} 
	 \label{Table:example}
\begin{adjustbox}{width=1\textwidth,center}
\begin{tabular}{llLllll}\hline
\toprule
\bfseries Category & \bfseries Sub-category & \bfseries Example (Excerpts from a related refactoring review discussion) \\
\midrule
\multirow{12}{*} {\textbf{Quality}} & \cellcolor{gray!30} Code smell & \cellcolor{gray!30} \say{\textit{I do understand the desire to refactor some code to \textbf{eliminate duplicate code}.  The purpose of the common class was to contain all of the duplicate code between the 2 drivers.   This seems like a half baked approach to refactoring to accomplish that goal, when the common class should have been used as a new base.  Now with this patch there are 2 classes (base and common) that contain common code.}} \\ 
& Internal quality attribute & \say{\textit{I think there is still an \textbf{inheritence problem} with the Base2 class (see inline) and the way python handles multiple inheritence means we have to be careful when creating the UTF8 versions of the unittest classes.}}\\
& \cellcolor{gray!30} External quality attribute & \cellcolor{gray!30} \say{\textit{Looking through the series i found enough such code to see that it does \textbf{not look easy readable}.}}\\
& Technical debt & \say{\textit{I was trying to use exception.ARQInvalidState with assertRaises, but \textbf{found we had some technical debts}, so use Exception instead to check the error message.}}\\ 
& \cellcolor{gray!30} Design pattern & \cellcolor{gray!30} \say{\textit{I think the alternative should be what other refactor possibilities are. The proposed change indicates some of the methods that are being planned to be implemented. As an alternative to evolve this library in future \textbf{we need another design pattern}.}}\\
& Coding convention & \say{\textit{We've seen how \textbf{difficult} it is to arrive at \textbf{a single naming convention} that works for all OpenStack projects, but as Rally expands beyond benchmarking OpenStack alone, it is quite simply impossible to create a single name format that will work for literally everything.}}\\ 
& \cellcolor{gray!30} Lack of documentation & \cellcolor{gray!30} \say{\textit{\textbf{Is there a documentation} somewhere explaining how to define "extra" dependencies? It would be nice to have a link somewhere. IHMO dependencies are a very complex problem, and extra dependencies look even more complex to me.}}\\
\hline
\multirow{9}{*} {\textbf{Refactoring}} & Refactoring correctness& \say{\textit{I think, this piece is \textbf{incorrectly refactored}. You should create neutron.conf.agent.linux with IP\_LIB for linux and corresponding change for neutron.conf.agent.windows with IP\_LIB for windows. Then you could import it, similar to this, how it was done previously.}} \\ 
& \cellcolor{gray!30} Behavior preservation & \cellcolor{gray!30} \say{\textit{Refactoring involves "behavior-preserving transformations", but \textbf{this change does change behavior}. Describing it as refactoring is misleading and suggests the change is less risky than it really is}}\\
& \begin{tabular}[c]{@{}l@{}}Refactoring co-occurrences \\with changes \end{tabular}  & \say{\textit{I would have \textbf{separate this change from your refactor}. The value returned by this function is transferred to the compute manager and to the virt drivers. Some of them could have considered True/False/None to handle different behavior and even if that is not the case, to have this isolated can help reviewer to ensure all is OK.}}\\
& \cellcolor{gray!30} Domain constraint & \cellcolor{gray!30} \say{\textit{Looking through the series i found enough such code to see that it does not look easy readable. It operates with StoragePolicy's properties which are initialized there and here we can guess only what they mean (e.g. external\_boot). \textbf{StoragePolicy knows all about disks, what kinds of disks exists, what disks are used by instance}, etc. It would be consistent to move such operations into that class as well. Or at least to add an iterator over disks there and to use it}} \\ \hline
\multirow{11}{*} {\textbf{Objective}} & Unclear goal & \say{\textit{with regards to the refactor. The plan was to make a progressive refactor that would keep the old API around while the new one is adopted. The \textbf{motivations behind this refactor} are related to the lack of any kind of architecture and design in the current code. The current code, as it is, is not easily consumable by other services, which is the whole point of this library.}} \\ 
& \cellcolor{gray!30} Unknown benefit & \cellcolor{gray!30} \say{\textit{I would like understand \textbf{what is the added value} or improvement with these changes to l3-agent scheduler.}} \\
& Potential side effect & \say{\textit{It seems promising, but the change in pattern, if not thoroughly thought out, may \textbf{lead to other unforeseen side effects}.}} \\
& \cellcolor{gray!30} Change scope & \cellcolor{gray!30} \say{\textit{Well I think IloISCSIDeploy name wins over IloPXEDeploy. This is no longer tied to a particular boot interface and can work with other boot interfaces (like virtual media when other driver is refactored).  But I missed what Ruby pointed out, it's doing pxe.<something> for now.  Needs to be refactored but may be better in another patch.  We have enough content in this patch alone which refactors the generic upstream driver.  So I think it's better to refactor in next patch.}}\\ 
& Feature support & \say{\textit{Isn't this a whole \textbf{new feature}, and not just a refactor? IMHO this should be its own spec.[...] Yes, this is new feature and cannot be implemented by only this refactoring. This feature depends on the refactoring. I agree with you to create a spec}}\\ 
& \cellcolor{gray!30} Bug fixing & \cellcolor{gray!30} \say{\textit{I don't think some of this patch is needed since we merged Gage's fix. On the other hand, this change seems to be doing quite a bit of refactoring. Is there a reason to continue pursuing that refactor, or was it mainly to \textbf{fix the bug?}}}\\ 
\hline
\multirow{6}{*} {\textbf{Testing}} & Lack of coverage & \say{\textit{\textbf{Is there enough coverage} on this to ensure that all changes are tested?}} \\
& \cellcolor{gray!30} Absence of result & \cellcolor{gray!30} \say{\textit{I should read more closely... this was intended according to the commit message: "for patches that get 'rechecked' a lot,
you'll see the whole \textbf{history of test results} on current patch".}}\\
& Poor test quality & \say{\textit{What do you think of continuing to clean up the test infrastructure as you've started here and create good examples of what to do?  We could document \textbf{how to write good tests} so that authors and reviewers could ensure the quality of new additions, and require rewriting of existing tests only as needed rather than attempting the monumental effort of fixing the entire unit test tree.}}\\
\hline
\multirow{7}{*} {\textbf{Integration}} & \cellcolor{gray!30} Configuration issue & 
\cellcolor{gray!30} \say{\textit{IMO also some detailed configuration example and illustration is needed in etc/designate.conf - My concern is that, not specifically against patch but for whole project, we add all these new functionalities but people don't know how to use/configurate it. An example is option 'options' in many config sections. It varies a lot from different types of backend/pool\_target, but we didn't document them well. People(at least me) need to read the source code to figure out \textbf{how to configure that correctly}. Maybe we should come up with a patch fix this in short future.}} \\    
& Merge conflict &  \say{\textit{This seems like it's mostly just going to \textbf{cause merge conflicts} with a lot of outstanding code.}} \\
& \cellcolor{gray!30} API management& \cellcolor{gray!30} \say{\textit{This \textbf{seems to be a breaking change} in the interface, meaning this new cli wouldn't work with the old api and vice versa. I think it would be appropriate if this new code (in the API) produced dag\_execution\_date on the action dictionary returned, so as to be backwards compatible with old clients, and this new client should probably sniff this for the new way, and revert to the old way if needed}} \\  
& Build failure & \say{\textit{\textbf{Build failed} (check pipeline).  For information on how to proceed, see http://docs.openstack.org/infra/manual/developers.html\#automated-testing.}} \\ 
\hline
\multirow{7}{*} {\textbf{Management}} & \cellcolor{gray!30} No review activity & \cellcolor{gray!30} \say{\textit{This patch has been \textbf{idle for a long time}, so I am abandoning it to keep the review clean sane. If you're interested in still working on this patch, then please unabandon it and upload a new patchset.}} \\
& Forgotten review & \say{\textit{I see that this change has not been updated since April 23rd.  I also see that there have been \textbf{no responses} from you to the questions Mark asked inline on the most recent patchset.  Do you intend to continue working on this change?}}\\
& \cellcolor{gray!30} Change dependency& \cellcolor{gray!30} \say{\textit{This is a significant enough change that I think it warrants \textbf{discussion with the drivers team and a spec}.}}\\
& Review guideline violation & \say{\textit{This is much more readable than before, but I would like to see a satisfactory answer to the question of why this is necessary, before this is merged. As far as I am aware, there is \textbf{no consensus} from OpenStack on \textbf{following this guideline}; do you have a spec for this work?}}\\ 

\bottomrule
\end{tabular}
\end{adjustbox}
\vspace{-.3cm}
\end{table*}
\noindent\textbf{Category \#6: Management.} Another category emerged from the manual coding analysis is review management. 
 The subcategories we found were \textit{no ongoing review activity}, \textit{forgotten review}, \textit{change dependency}, and \textit{review guideline violation}. In other words,  we observe that some of the reviews take longer due to a lack of reviewer attention, as there is a case of little prompt discussion about the proposed changes. For instance, some reviews received a review score of +1 from a reviewer, but there was no other activity after waiting a couple of days or months. \textcolor{black}{The follow-up interview confirms that this category is common in industry. Reviewers have their own workload, and some discussions can be subject to delays depending on the developer and reviewers' participation \cite{thongtanunam2015investigating}.} 

\section{Implications}
\label{Section:Implications}

\subsection{Implications for Practitioners}
\textbf{Establishing guidelines for refactoring-related reviews.} Our taxonomy shows reviewing refactoring goes beyond improving the code structure. To improve the practice of reviewing refactored code, and contribute to the quality of reviewing code in general, managers can collaboratively work with developers to establish customized guidelines for reviewing refactoring changes which could establish beneficial and long-lasting habits or themes to accelerate the process of reviewing refactoring. Additionally, since our RQ2 findings show that integration and testing are one of the challenges caught by developers when reviewing refactoring changes, it is recommended to utilize continuous integration to keep the testing suite in sync with the code base during and after refactoring. Further, adherence to coding conventions is considered one of difficulties when reviewing refactoring tasks. To cope with this challenge, we recommend project leaders to educate their developers about the coding conventions adopted in their systems. Considering the above-mentioned characteristics not only save developers time and effort, but also can assist taking informed decision and bring a discipline toward reviewing code involving refactorings within a software development team.



\subsection{Implications for Researchers}
\textbf{Exploring the potential of combining multiple behavior preservation strategies when reviewing refactorings.} Our study shows that preserving the behavior of software refactoring during the code review process is a critical concern for developers, and developers determine whether a behavior is preserved based on the context.  Recently, AlOmar \etal \cite{alomar2021preserving} aggregated the behavior preservation strategies that have been evaluated using single or multiple refactoring operations, and some of these refactorings are applied using multiple
strategies. In order to accelerate the process of reviewing code involving refactoring, future researchers are advised to explore the potential of combining several behavior preservation approaches and use those  which would be useful in the context of modern code review according to a defined set of criteria. \textcolor{black}{For instance, Soares \etal \cite{soares2009saferefactor} have implemented the tool `SafeRefactor' to identify behavioral changes in transformation. It would be an interesting idea to verify the correctness after the application of refactoring by embedding test results and generating a test suite for capturing unexpected behavioral changes in code review board.}

\textbf{Supporting for the refactoring of non-source code artifacts.} From RQ2, we discover that refactoring operations are not limited to source code files. Artifacts such as databases and log files are also susceptible to refactoring. Similarly, we also observed discussions about refactoring test files. While it can be argued that test suites are source code files,  recent studies by \cite{pantiuchina2020developers,alomar2021ESWA} show that the types of refactoring operations applied to test files are frequently different from those applied to production files. Hence, future research on refactoring is encouraged to introduce refactoring mechanisms and techniques exclusively geared to refactoring non source code artifact types and test suites.


\subsection{Implications for Tool Builders}
\textbf{Developing next generation refactoring-related code review tools.}  Finding that reviewing refactoring changes takes longer than non-refactoring changes reaffirms the necessity of developing accurate and efficient tools and techniques that can assist developers in the review process in the presence of refactorings. Refactoring toolset should be treated in the same way as CI/CD tool set and integrated into the tool-chain. 
 Researchers could use our findings with other empirical investigations of refactoring to define, validate, and develop a scheme to build automated assistance for reviewing refactoring considering the refactoring review criteria as review code become an easier process if the code review dashboard augmented with the factors to offer suggestions to better document the review. 
  
 Moreover, we noticed that poorly naming the code elements is one of the major bad refactoring practices typically catch by developers when reviewing refactoring changes. Constructing tools that enforce code conventions aid in speeding up reviewing refactoring changes and benefiting reviewers to focus on deeper design defects. \textcolor{black}{Furthermore, to accelerate code review process and limit having a back-and-forth discussion for clarity on the problem faced by the developer, tool builders can develop \textit{bots} for the integration, testing, and management categories. Additionally, it would be interesting to use a popular and widely adopted quality framework, \eg,  Quality Gate of SonarQube \cite{olivier2013sonar}, as part of quality verification process by embedding its results in the code review. This might facilitate convincing the reviewer about the impact and the correctness of the performed refactoring.}


\section{Threats To Validity}
\label{Section:Threats}

In this section, we describe potential threats to validity of our research method, and the actions we took to mitigate them.

\textbf{Internal Validity.} Concerning the identification of refactoring related code review, we select reviews with the keyword `\textit{refactor*}' in their title and description. Such selection criteria may have resulted in missing refactoring-related reviews and there is the possibility that we may have excluded synonymous terms/phrases. However, even though this approach reduces the number of reviews in our dataset, it also decreases the false positiveness of our selection. While our data collection may results in missing some reviews, our approach ensures that we analyze reviews that are explicitly geared towards refactoring. In other words, these are reviews where developers were explicitly documenting a refactoring action and they wanted it to be reviewed. Additionally, upon performing the manual inspection on review discussions, we realized that refactoring is heavily emphasized on discussions that start with a title or a description containing the keyword `\textit{refactor*}'. Yet, this does not prevent other discussions from bringing refactoring into the picture, and these will be missed by our selection (\ie, false negatives). We opted for such picky selection to only consider discussions when code authors explicitly wanted their refactored code to be reviewed, and so reviewers eventually propose a refactoring-aware feedback, which is what we are aiming for in this study. Therefore, it would be interesting to consider scenarios where reviewers have raised concerns about refactoring a code change that was not intended to be associated with refactoring. Since refactoring can easily be interleaved with other functional changes, it would be interesting to extract scenarios where reviewers thought it was misused. Study can also help developers better understand not only how to refactor their code, but also how to document it properly for easier review.

Further, we focus on the code review activity that is reported by the tool-based code review process, \ie, Gerrit, of the studied systems due to the fact that other communication media (\eg, in-person discussion \cite{beller2014modern}, a group IRC \cite{shihab2009studying}, or a mailing list \cite{rigby2011understanding}) do not have explicit links of code changes and recovering these links is a daunting task \cite{thongtanunam2016revisiting,bacchelli2010linking}.

\textbf{Construct Validity.} About the representativeness and the correctness of our refactoring review criteria, we derive these criteria from a manual analysis of a subset of refactoring-related reviews that have lengthier review duration. This approach may not cover the whole spectrum of all the review criteria done with refactoring in mind. To mitigate this threat, we reviewed the top 7.3\% lengthier refactoring reviews assuming that these reviews will capture the most critical challenges. Additionally, to avoid personal bias during the manual analysis, each step in the manual analysis was conducted by two authors, and the results were always cross-validated. Another potential threat
to validity relates to refactoring reviews. Since refactorings could interleave with other changes \cite{murphy2012we} (\ie, developers performed changes together with refactorings), we cannot claim that the selected refactoring reviews are exclusively about refactoring. Nevertheless, during our qualitative analysis, we identified this activity as one of the challenges that contribute to slowing down the review process.

\textbf{External Validity.} We focus our study on one open-source system, due to the low number of systems that satisfied our eligibility criteria (see Section \ref{Section:Methodology}). Therefore, our results may not generalize to all other open-source systems or to commercially developed projects. However, the goal of this paper is not to build a theory that applies to all systems, but rather to show that refactoring can have an impact on code review process. Another potential threat relates to the proposed taxonomy. Our taxonomy may not generalize to other open source or commercial projects since the refactoring review criteria may be different for another set of projects (\eg, outside the OpenStack community). Consequently, we cannot claim that the results of refactoring review criteria (see Figure \ref{fig:challenges}) can be generalized to other software systems where the need for improving the design might be less important. \textcolor{black}{To mitigate this threat, we validate the taxonomy with an experienced software developer, by conducting a follow-up interview to gather further insight and possible clarification. \textcolor{black}{Yet, performing the validation with only one developer brings its own bias. The choice of one developer was driven by their experience with code review. We mitigated the bias by selecting a developer that does not belong to the software systems we analyzed. This brings an external opinion that has no conflict of interests with the current projects.}}

\textbf{Conclusion Validity.} To compare between two groups of code review requests, we used appropriate statistical procedures with \textit{p}-value and effect size measures to test the significance of the differences and their magnitude.  A statistical test was deployed to measure the significance of the observed differences between group values. This test makes no assumption that the data is normally distributed. Also, it assumes the independence of the groups under comparison. We cannot verify whether code review requests are completely independent, as some can be re-opened, or one large code change can be treated using several requests. To mitigate this, we verified all the reviews we sampled for the test.

\section{Conclusion}
\label{Section:Conclusion}

Understanding the practice of refactoring code review is
of paramount importance to the research community and industry. Although modern code review is widely adopted in open-source and industrial projects, the relationship between code review and refactoring practice remains largely unexplored. In this study, we performed a quantitative and qualitative study to investigate the review criteria discussed by developers when reviewing refactorings. Our results reveal that reviewing refactoring changes take longer to be completed compared to non-refactoring changes, and developers rely on a set of criteria to develop a decision about accepting or rejecting a submitted refactoring change, which makes this process to be challenging. 

For future work, we plan on conducting a structured survey with software developers from both open-source and industry. The survey will explore their general and specific review criteria when performing refactoring activities in code review. This survey will complement and validate our current study to provide the software engineering community with a more comprehensive view of refactoring practices in the context of modern code review. Another interesting research direction is to link refactoring-related reviews to refactoring detection tools such as the Refactoring Miner \cite{tsantalis2018accurate} or RefDiff \cite{silva2017refdiff} to better understand the impact of these reviews on refactoring types specifically.

\bibliographystyle{ACM-Reference-Format}
\bibliography{sample-base}

\end{document}